\documentclass[twocolumn,trackchanges]{aastex61}

\usepackage{rotating,epsfig}

\def\msun{\rm M_{\sun}}

\begin{document}


\shortauthors{Reipurth et al.}
\shorttitle{The Haro 5-2 Quadruple System}

\title{Haro 5-2: A New Pre-Main Sequence Quadruple Stellar System}

\author[0000-0001-8174-1932]{Bo Reipurth}
\affiliation{Institute for Astronomy, University of Hawaii at Manoa,
640 N. Aohoku Place, HI 96720}
\affiliation{Planetary Science Institute, 1700 E Fort Lowell Rd, Suite 106, Tucson, AZ 85719}

\author[0000-0001-7124-4094]{C. Brice\~{n}o}
\affiliation{NSF's NOIRLab/Cerro Tololo Inter-American Observatory, Casilla 603, La Serena, Chile}

\author[0000-0003-2824-3875]{T. R. Geballe}
\affiliation{Gemini Observatory, 670 North Aohoku Place, Hilo, HI 96720}

\author[0000-0002-1917-9157]{C. Baranec}
\affiliation{Institute for Astronomy, University of Hawaii at Manoa,
640 N. Aohoku Place, HI 96720}

\author[0000-0003-1448-8767]{S. Mikkola}
\affiliation{Department of Physics and Astronomy, University of Turku, Yliopistonmki (Vesilinnantie 5), Finland}

\author[0000-0002-3656-6706]{A.M. Cody}
\affiliation{SETI Institute, 339 N Bernardo Ave, Suite 200, Mountain View, CA 94043}

\author[0000-0002-8293-1428]{M. S. Connelley}
\affiliation{Institute for Astronomy, University of Hawaii at Manoa,
640 N. Aohoku Place, HI 96720}
\affiliation{Staff Astronomer at the Infrared Telescope Facility, which is operated by the University of Hawaii under contract NNH14CK55B with the National Aeronautics and Space Administration}

\author[0000-0002-8591-472X]{C. Flores}
\affiliation{Institute of Astronomy and Astrophysics, Academia Sinica, 11F of AS/NTU Astronomy-Mathematics Building, No.1, Sec. 4, Roosevelt Rd, Taipei 10617, Taiwan, R.O.C. }

\author[0000-0001-5306-6220]{B. A. Skiff}
\affiliation{Lowell Observatory, 1400 West Mars Hill Road, Flagstaff, AZ 86001}

\author{J. D. Armstrong}
\affiliation{Institute for Astronomy, University of Hawaii at Manoa, 34 Ohia Ku St., Pukalani, HI 96768}

\author[0000-0001-9380-6457]{N. M. Law}
\affiliation{Department of Physics and Astronomy, University of North Carolina at Chapel Hill, Chapel Hill, NC 27599-3255}

\author[0000-0002-0387-370X]{R. Riddle}
\affiliation{Cahill Center for Astrophysics, California Institute of Technology, 1216 East California Boulevard, Pasadena, CA 91125}

\email{reipurth@hawaii.edu}
\correspondingauthor{Bo Reipurth}

\begin{abstract}


We have discovered that the H$\alpha$ emission line star Haro~5-2,
located in the 3-6~Myr old Ori~OB1b association, is a young quadruple
system. The system has a 2+2 configuration with an outer separation of
2.6 arcseconds and with resolved subarcsecond inner binary components. 
The brightest component, Aa, dominates the A-binary, it is a weakline
T~Tauri star with spectral type M2.5$\pm$1. The two stars of the B
component are equally bright at J, but the Bb star is much
redder. Optical spectroscopy of the combined B pair indicates a rich
emission line spectrum with a M3$\pm$1 spectral type. The spectrum is
highly variable and switches back and forth between a classical and a
weakline T~Tauri star. In the near-infrared, the spectrum shows
Paschen~$\beta$ and Brackett~$\gamma$ in emission, indicative of
active accretion. A significant mid-infrared excess reveals the
presence of circumstellar or circumbinary material in the system.
Most multiple systems are likely formed during the protostellar phase,
involving flybys of neighboring stars followed by an in-spiraling
phase driven by accretion from circumbinary material and leading to
compact sub-systems. However, Haro 5-2 stands out among young 2+2
quadruples as the two inner binaries are unusually wide relative to
the separation of the A and B pair, allowing future studies of the
individual components. Assuming the components are coeval, the system
could potentially allow stringent tests of PMS evolutionary models.



\end{abstract}



\section{Introduction}
\label{sec:introduction}

The majority of stars are born as part of
binary or multiple systems; for reviews see \cite{duchene_kraus2013},
\cite{reipurth2014}, and \cite{offner2022}. It has even been
suggested that {\em all} stars are born in binaries or multiples, and
that the distribution of singles, binaries, triples, etc. observed in
the field is the result of subsequent dynamical interactions
\citep{larson1972}. This is consistent with simulations
\citep{delgado-donate2004} and observations, which show an excess of
binaries among T~Tauri stars compared to the field
\citep[e.g.][]{reipurth1993} and an even higher multiplicity fraction among
embedded protostars \citep{connelley2008,chen2013}.

While triple stellar systems have been studied extensively, both
observationally as well as numerically, quadruple systems have
received much less attention. 

It is well established that a non-hierarchical quadruple system is
unstable, and will quickly transform into a hierarchical either 2+1+1
or 2+2 configuration or break up by ejecting a single or binary star.

Many quadruple systems are known, most of them found in recent years
through major sky surveys 
\citep[e.g.][]{kostov2022, kostov2024}.
Most famous of all is $\epsilon$~Lyrae, a system of 4 similar A-type
stars, which was first noticed by William Herschel\footnote{Herschel
writes about $\epsilon$~Lyrae on August 29, 1778: {\em $''$A very
curious double-double star. At first sight it appears double at some
considerable distance, and by attending a little we see that each of
the stars is a very delicate double star$''$} - Pratt \& Gledhill
(1880)}.

Detailed observational studies of quadruple systems have until
recently been limited, reflecting the rarity of such systems. With the
advent of deep all-sky surveys, the number of quadruple systems is
rapidly increasing, e.g. \citet{zasche2019}, \citet{fezenko2022}. Although such surveys have inherent biases, they
still offer new insights into the properties of quadruple systems, in
particular the double eclipsing systems \citep{vaessen2024}.

A statistical study of multiplicity in F-G stars was
performed by \cite{tokovinin2014}, who finds the following percentages for
singles, binaries, triples, and quadruples: 54:33:8:4. Among these
quadruples, the number of 2+2 systems is higher than the 2+1+1
systems. \cite{raghavan2010} and \cite{riddle2015} also found a
preponderance of 2+2 systems.

An understanding of this difference must be found in the early stages
of evolution of quadruple systems shortly after they are formed, and
hence the study of newborn quadruple systems is important. Only a few 2+2 quadruple
pre-main sequence systems are known, among others HD~98800 
\cite[e.g.][]{prato2001,zuniga-fernandez2021}, FV~Tau and J4872 \citep{correia2006}, 
LkCa~3 \citep{torres2013}, 2M0441+2301 \citep{todorov2010,bowler2015}, 
EPIC 203868608 \citep{wang2018}, and especially GG~Tau \citep[e.g.][]{white1999,dutrey2016}, 
with which Haro~5-2 has a remarkable similarity in terms of configuration. 
Most recently, TIC 278956474 was
found with the TESS mission to consist of two low-mass short-period
eclipsing binaries, with an age of 10-50 Myr \citep{rowden2020}.

We here
present the discovery of a new 2+2 pre-main sequence system known as  
Haro~5-2, located at  5:35:07.5, -2:49:00 (2000) in the Ori~OB1b 
association \citep{haro1953}. We provide detailed optical/infrared
imaging and spectroscopy, and discuss the properties and possible
formation scenarios of this system.

\section{Observations}
\label{sec:observations}

One of us (BAS) identified Haro~5-2 as a binary and it was included as
SKF2259 in The Washington Visual Double Star Catalog
\citep{mason2022}. It was subsequently observed on UT 2014 November 9
using the Robo-AO autonomous laser adaptive optics system at the
Palomar 1.5m telescope \citep{baranec2014}.  We performed the
observations using a long-pass filter with a cut-on wavelength of 600
nm and with a total exposure time of 3 minutes. The Ba-Bb pair is well
resolved. The Aa component has an approximate image width of
0$\farcs$15, with a non-resolved elongation in the direction of
Ab. The elongation was not present in the images of any other stars in
the field, so we suspected this was due to an unresolved companion. To
investigate this possibility, we observed Haro 5-2 with the NIRC2
instrument behind the Keck II adaptive optics system on UT 2015 August
5. We obtained six 30s exposures with the Kp filter and then five 30s
exposures with the H filter. All four components of the Haro 5-2
system are well resolved in the Kp images, while the seeing degraded
rapidly during the H-band observations making it difficult to resolve
the Aa-Ab pair.  More detailed observations are required
to search for additional components, as found in the similar GG~Tau
system \citep{difolco2014}.

\begin{figure*}
\centerline{\includegraphics[angle=0,width=12cm]{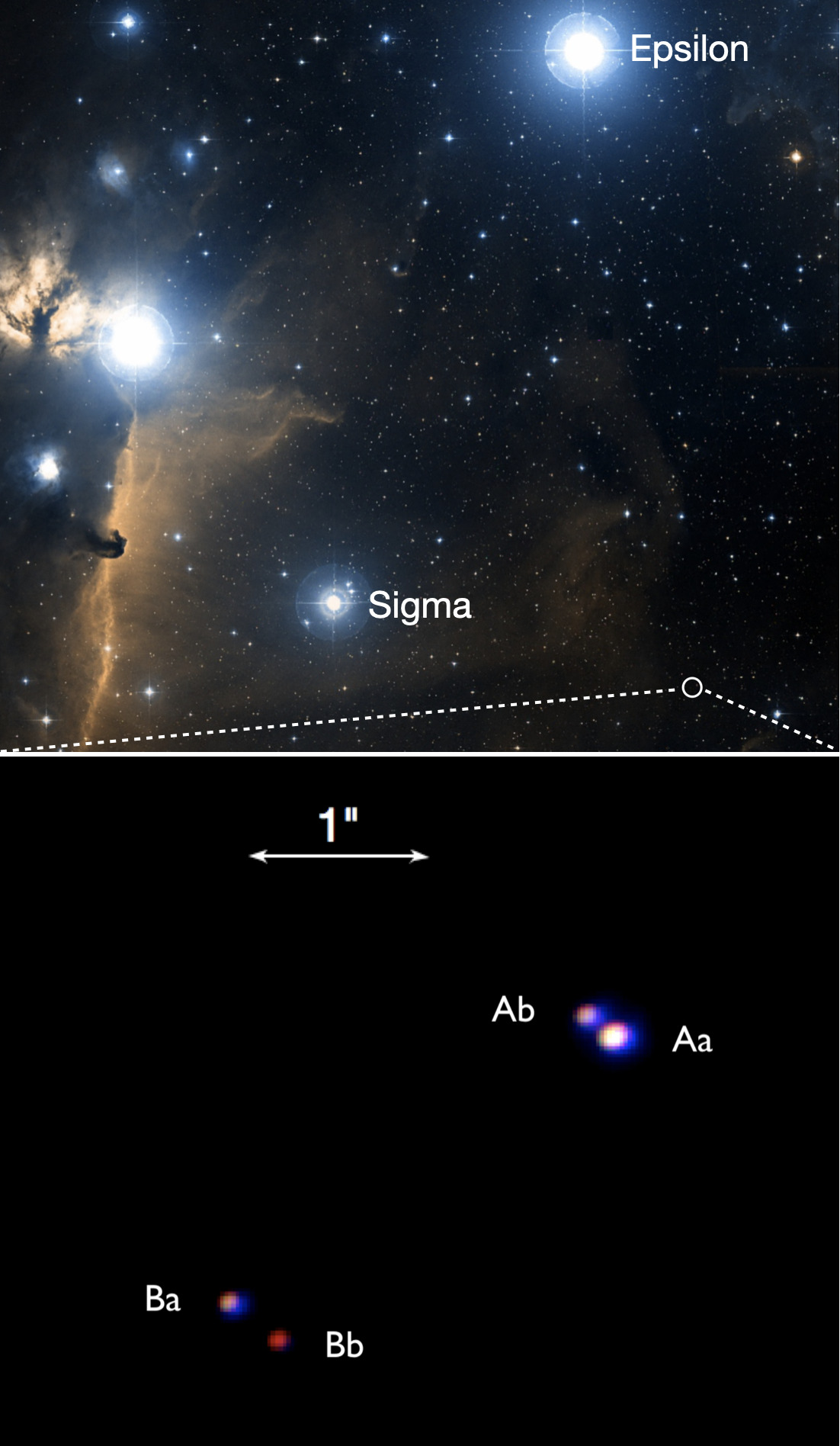}}
\caption{(top) Haro~5-2 is located in a little-studied part of Orion west of $\sigma$~Ori and south of $\epsilon$~Ori, both of which are associated with groups of low-mass young stars.  Image from ESASky.
(bottom) Near-infrared JHK color composite of Haro~5-2 obtained with adaptive optics on the Gemini-North telescope. The small blue halos are due to a slight elongation of the J image.
\label{location}}
\end{figure*}

Direct images of Haro~5-2 were obtained on UT 2015 August 27 at the
Gemini-North telescope with NIRI \citep{hodapp2003}  
at f/32 (0.02 arcsec pixels) and adaptive optics, 
see Figure~\ref{location}. In the J- and
H-bands a 3$\times$3 mosaic was obtained with individual 3-second
exposures, in total 81~sec, while in the K-band the exposures totaled
162~sec.

Haro~5-2 was observed with GNIRS \citep{elias2006} 
at the Gemini-North telescope on UT
2015 September 15 in a seeing of 0$\farcs$5, using 0.05 arcsec pixels 
and the 10 l/mm grating in cross-dispersed mode.
A 0$\farcs$15 wide slit
with a length of 5$\farcs$1 was placed along the Aa-Ab pair at position
angle 47$^\circ$. 12 exposures each of 60 seconds were acquired while
nodding $\pm$1$\farcs$5 along the slit. The pair was not resolved. The
slit was then changed to a position angle of 54.3$^\circ$ and placed
along the Ba-Bb pair. 12 exposures each of 120 seconds were acquired.
The two components could be resolved at H and
K, but not at J, and separate spectra were extracted in H and K. 
The unresolved Aa-Ab pair was reduced with the Gemini pipeline, while
the Ba-Bb pair was reduced by using IRAF \citep{tody1986,tody1993} 
and Figaro \citep{currie2014}.

We obtained spectroscopy of the Haro A and B components,
separated by 2$\farcs$62, with the Goodman High Throughput
Spectrograph \citep[GHTS -][]{clemens2004}, installed on the SOAR 4.1m
telescope on Cerro Pach\'on, Chile.  We used available time slots
during the engineering nights of Oct 19 and Nov 18, 2021, Nov 9, 2022, 
Jan 6, 2023 and Mar 9, 2023, for a total time span of about 1.4~years.  
The GHTS is a highly configurable imaging spectrograph that employs
all-transmissive optics and Volume Phase Holographic Gratings,
resulting in high throughput for low to moderate resolution
spectroscopy over the 320-850 nm wavelength range.  We used the
Goodman RED camera, which uses a deep-depletion CCD that provides
extended sensitivity into the red end of the spectrum with minimal
fringing.  The spatial scale of the Goodman detector is
0$\farcs15$/pixel, 
and the slit was set at a position angle of 214.7$^\circ$, with the
Atmospheric Dispersion Corrector active.

We configured the spectrograph for both low- and high-resolution
setups. For the low resolution setup we used the 400 l/mm grating in
its 400M1, and 400M2 $+$ GG 455 filter, preset modes, with 2x2
binning.  These two configurations combined span the wavelength range
$\sim 3600 \la \, \lambda \, \la 9000~$\AA. We used the $1\arcsec$
wide long slit, which yields a FWHM resolution $=6.7$~{\AA }
(equivalent to $\rm R=830$). We used exposure times of 120s for the
brighter Haro 5-2A and 300s for the fainter Haro 5-2B.  A total of 21
low-resolution spectra were obtained for Haro 5-2A (13 in the 400M1
setup and 8 with the 400M2 setup), while 25 spectra were obtained for
Haro 5-2B (13 with the 400M1 setup and 12 with the 400M2 one), during
a time window of nearly 506 days or about 1.4 years, from UT
2021-10-20 to UT 2023-03-10 (Table \ref{soar_ghts_obs}).

For the higher resolution spectra we used the 2100 l/mm grating
centered at 650~nm, with the 0$\farcs$45 long slit, and 1x2
binning. This setup produces a FWHM resolution of $0.45$~{\AA }
(or $\rm R=11,930$), 
spanning $565$~{\AA }, from $6140$~{\AA } to $6705$~{\AA}. 
Our choice of $\times 2$ binning in the spatial direction improved 
the signal-to-noise ratio 
while still matching the median 0$\farcs$7 seeing at SOAR. We
obtained $3\times 900$s exposures for the A component and $3\times
1200$s for the B component. We took a total of 6 higher resolution
spectra for each of the A and B components, 3 each on Oct 21, 2021 and 3
each on Nov 19, 2021 (Table \ref{soar_ghts_obs}).
The basic CCD reduction, up to and including the extraction of the 1-D
spectra, was carried out using the Goodman Spectroscopic Pipeline
\citep{torres-robledo2020}.  Wavelength calibration and final
combination of the spectra was done with IRAF.  We did not perform
flux calibration, since the main purpose of our follow up spectroscopy
was to characterize each component: determining its spectral type,
measuring line equivalent widths and determine its accretion
status. We measured equivalent widths of spectral features using the
{\sl splot} routine in IRAF.  The S/N ratio of the individual spectra
was typically $\sim 20-25$ at H$\alpha$, sufficient for measuring
equivalent widths down to $\sim 0.05$~{\AA }, at our highest spectral
resolution of $\sim 0.45$~{\AA } FWHM. In Figures \ref{soar_lowres}
and \ref{soar_hires} we show the combined spectra of Haro 5-2 obtained
with the low resolution, and the higher resolution setups,
respectively.  We measured equivalent widths in each of the individual
spectra. The uncertainty values quoted in Tables
\ref{soar_spec_lowres} and
\ref{soar_spec_hires} are the actual dispersion in measurement
values.


On the same nights that we obtained optical spectra on SOAR, we also
obtained near-IR spectroscopy resolving the Haro 5-2A and B
components using the TripleSpec 4.1 (TSpec) near-IR spectrograph
\citep{schlawin2014}.  
The fixed-format slit is 1$\farcs$1 wide and $28\arcsec$ long. The
spectral resolution is $\rm R\sim 3500$ across the six science orders.
We obtained three ABBA sequences for each A and B component, using 
exposure times of 30s for both components, with Fowler Sampling 4. 
The telluric standard HIP 26812 was observed contiguous with Haro
5-2, a single ABBA sequence was obtained at the same airmass, with an
exposure time of 5s, 2 coadds, and Fowler Sampling 1.
The slit position angle was set at 214.7$^\circ$.
The data were reduced in the standard way using the customized version
of the IDL Spextool package \citep{cushing2004} that was modified by
Dr. Katelyn Allers for use with TSpec at SOAR.

Haro 5-2 was observed by the Transiting Exoplanet Survey Satellite
(TESS) in its sector 6 (11 December 2018 to 7 January 2019) and sector
32 (19 November 2020 to 17 December 2020)
campaigns, which lasted approximately 27 days each. Known as TIC
427380741 in the TESS Input Catalog, Haro 5-2 was monitored in full
frame image (``FFI") mode. Sector 6 images were acquired at a cadence
of 30 minutes, while those for sector 32 were taken every 10 minutes.
The TESS mission data products for FFI targets do not include light
curves, and so we downloaded stacked $15\times 15$ pixel cut-out
images in the form of target pixel files (TPFs) using the
\texttt{lightkurve} package \cite{lightkurve2018}. In
these images, Haro 5-2 appears clearly in the center, approximately
two pixels offset from the nearest relatively bright source. To create
a light curve, we placed a $3\times 3$ pixel square aperture on the
source, summing the flux at each available time. \texttt{lightkurve}
was used to carry out this procedure, as well as to subtract out a
star-free average background. TESS light curves of different targets
often exhibit common systematic trends due to scattered light, which
can be removed by detrending against flux time series of pixels
outside the target aperture. We employed the \texttt{lightkurve}
RegressionCorrector class to remove low-level trends in the Haro 5-2
light curves, using the top two principal component vectors.  We
ultimately removed points for which the original image TPF data was
flagged in the TESS pipeline as being contaminated by stray light. The
resulting light curves displayed a series of significant undulations
on a range of timescales, similar to what was seen in the raw time
series.

\section{Observational Results}
\label{sec:results}

\subsection{Haro~5-2}

Haro~5-2, also known as 2MASS J05350753-0248596 and ESO-H$\alpha$ 1108
\citep{pettersson2014}, is located in the Ori~OB1b association. As
discussed in the following, we conclude that it is a young stellar
object (YSO) based on its H$\alpha$ emission, its mid-infrared excess,
its irregular variability, and its location within a star forming
region.

In the first survey for H$\alpha$ emission stars in the $\sigma$~Ori
region, \cite{haro1953} discovered 98 objects, including
Haro~5-2, which was also detected in the H$\alpha$ emission line
survey by \cite{weaver2004}. The $\sigma$~Ori cluster has an
age of 3-5~Myr and contains several hundred young low mass stars
\citep[e.g.][]{caballero2008a,hernandez2014,koenig2015}.

Haro~5-2 is located in a little-studied region southwest of the
cluster of young stars surrounding the O9.5~V multiple star
$\sigma$~Ori. \cite{schaefer2016} measured the distance to
$\sigma$~Ori itself to be 387.5$\pm$1.3~pc, while \cite{caballero2018}
used Gaia~DR2 for all known members of the $\sigma$~Ori cluster to
derive a mean distance of 391$\pm$44~pc.

\cite{caballero2008b} stated that ``the cluster seems to have two
components: a dense core that extends from the centre to r$\sim$20
arcmin and a rarified halo at larger distances.'' This halo extends to
$\sim$30~arcmin, whereas Haro~5-2 has a separation to $\sigma$~Ori of
almost one degree. Consequently, Haro~5-2 may alternatively be part of
the loose clustering Collinder~70 surrounding $\epsilon$~Ori, also known as 
Alnilam \citep{collinder1931,caballero_solano2008}.  The $\sigma$~Ori
and Collinder~70 populations together form the subgroup Ori~OB1b
defined by \cite{blaauw1964}.  \cite{kounkel2018} suggest that Ori~OB1b
is located at a distance of 357$\pm$3~pc and has an age of 3-6
Myr. Figure~\ref{location}a shows the location of Haro~5-2 relative to
$\sigma$~Ori and $\epsilon$~Ori.

Haro 5-2 is not an isolated H$\alpha$ emission star, \cite{haro1953} 
found three other H$\alpha$ emitters, Haro 5-1, -3, and -4,
within a few arcminutes, and within 10 arcminutes \cite{pettersson2014} 
found another five: ESO-H$\alpha$ 1011, 1014, 1050, 1146, and
1364. A variable YSO, V2070~Ori, is also located close to
Haro~5-2. Using Gaia DR3 parallaxes for the stars in this little
group (omitting the multiple system Haro~5-2 itself and also Haro~5-3 for
neither of which Gaia has a parallax) suggests a mean distance 
of 373$\pm$13~pc, which we adopt.

Haro~5-2 is not directly associated with any cloud, but a small
cloudlet, Dobashi~4834, is located about 15~arcmin to the south-west
\citep{dobashi2011} and is part of the shell of gas and dust that has been
pushed away from the central O-star $\sigma$~Ori, see Figure~1 of
\cite{koenig2015} which shows a WISE 3-color image mosaic of the
region. 

\begin{figure*}
\centerline{\includegraphics[angle=0,width=13.5cm]{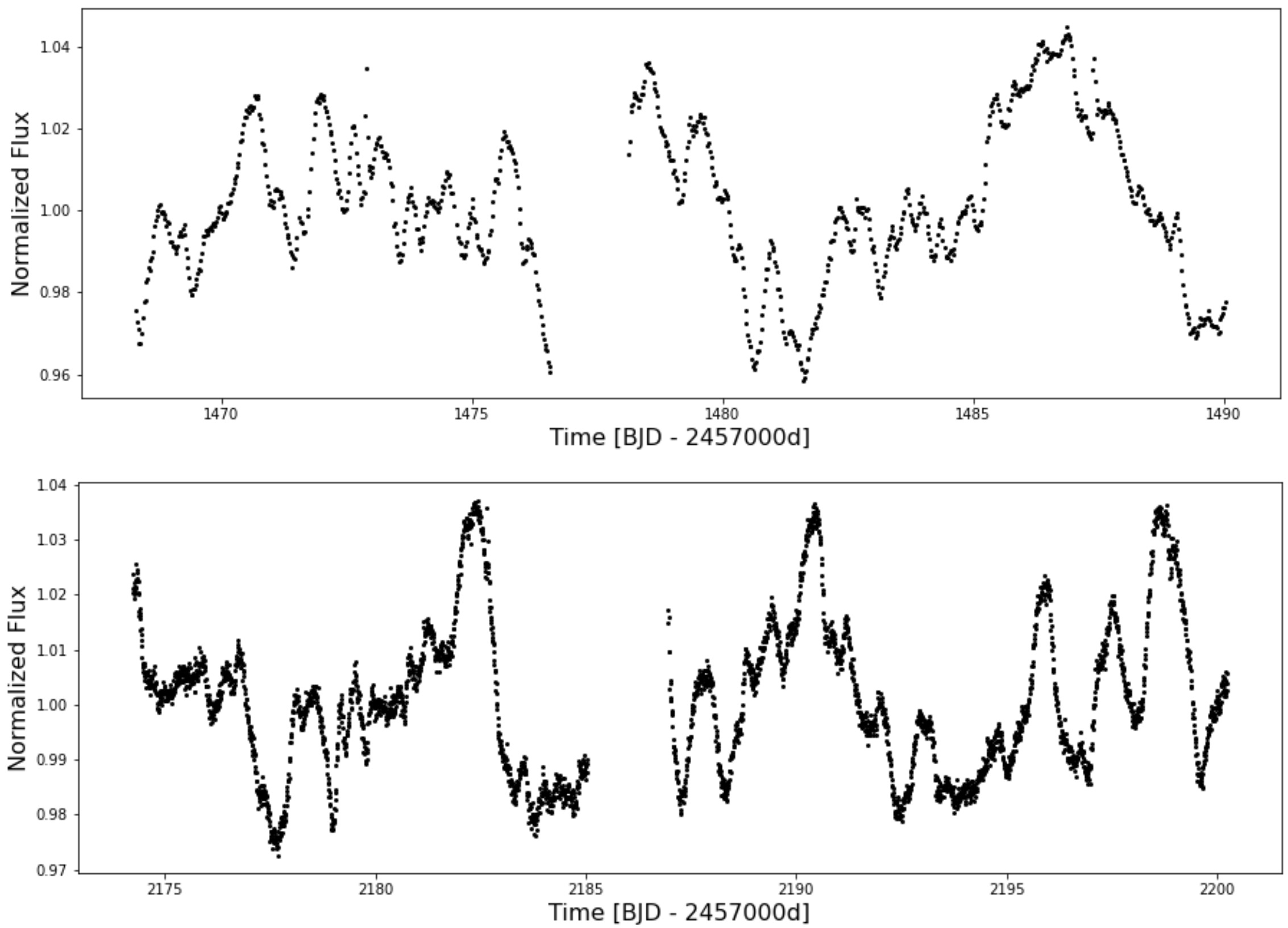}}
\caption{TESS light curves of the combined light from the four components of Haro 5-2  from around December 2018 (Sector 6, top) and
around December 2020 (Sector 32, bottom). For Sector 6 the cadence was
once every 30 minutes and for Sector 32 it was once every 10 minutes.
  \label{TESS}}
\end{figure*}

Given their youth, the components of Haro 5-2 are likely to be
variable, and a {\em g}-lightcurve from ASASSN
\citep{shappee2014, kochanek2017} of the integrated system light
indeed shows irregular variability over a 5-year period with
characteristic amplitudes of 0.1 - 0.2 magnitudes. TESS has observed the region including
Haro 5-2 on two occasions (see Section~\ref{sec:observations}) and the
light curves are shown in Figure~\ref{TESS}.  The light curves vary
with peak-to-peak amplitudes of $\sim$7\% and are largely of a
stochastic type that may indicate accretion variability
\citep{cody2014}. This irregular variability likely originates in the B-components, of which at least one is a classical T Tauri star (see
Section~\ref{subsubsec:haro5-2b}).  In addition, the data from sector
32 exhibit several upward excursions of a few percent amplitude that
are reminiscent of distinct accretion bursts
\citep{cody2017}.

Since we are viewing the combined light of several stars, it is
possible that multiple variability types are represented in the time
series. For each light curve, we computed a Fourier transform
periodogram and searched for persistent signals. We find one tentative
periodicity in common between both TESS light curves, at a timescale
of 1.33 days (sector 6) to 1.36 days (sector 32) and an amplitude of
$\sim$0.7\%. These detections are tentative since the light curves are
dominated by higher amplitude stochastic behavior; the signal-to-noise
ratio in the periodogram is 3.2 (sector 6) to 3.6 (sector 32), whereas
a secure detection would require a signal-to-noise value of at least
4.0 \citep[e.g.][]{breger1993}.  If real, the possible periodic signals could
be indicative of starspots on one member of Haro 5-2, probably
component Aa, which is significantly brighter than the other
stars.

\subsection{Infrared Imaging}

Figure~\ref{location}b shows a multi-color JHK image of Haro~5-2,
obtained with adaptive optics at the Gemini-N~8m telescope, in which
Haro~5-2 is well resolved as a 2+2 quadruple system.  What makes
Haro~5-2 of particular interest is that the two close pairs have
rather large projected separations Aa-Ab and Ba-Bb relative to the
projected separation of A-B compared to other PMS quadruple systems.
If not purely a projection effect, this might be due to its youth,
with the four components still interacting before settling into a
long-term stable configuration in which the inner binaries have
hardened.

We have measured the separations for the three pairs and determined
the following separations for Aa-Ba: 2$\farcs$61 (975~AU), Aa-Ab:
0$\farcs$19 (72~AU), and Ba-Bb 0$\farcs$35 (131~AU). Also, Andrei Tokovinin
kindly observed the system with the high resolution speckle camera at
the SOAR telescope \citep{tokovinin2022} and these values are listed in Table~\ref{tokovinin}.
The projected separation of the A and B binaries of about 1000~AU may suggest 
a period around 30,000~yr. The closest binary, Aa-Ab,  has a period of $\sim$1000~yr, 
which will be measurable in a time span of a few years.

Photometry of the components from the Gemini data on UT 2015 Aug 27 yields \\
\indent Aa: {\em K}=11.53  ~~~ {\em H-K}=0.40 \\
\indent Ab: {\em K}=12.34  ~~~ {\em H-K}=0.30 \\
\indent Ba: {\em K}=12.46  ~~~ {\em H-K}=0.34 \\
\indent Bb: {\em K}=12.54  ~~~ {\em H-K}=0.92 \\
The uncertainties in H and K are around 0.02 mag. 
While the three brighter components have roughly the same colors, Bb
is extremely red, suggesting that this component is either highly
extincted, has a strong near-infrared excess, or a mixture of the two.

\begin{figure*}
\centerline{\includegraphics[angle=0,width=17cm]{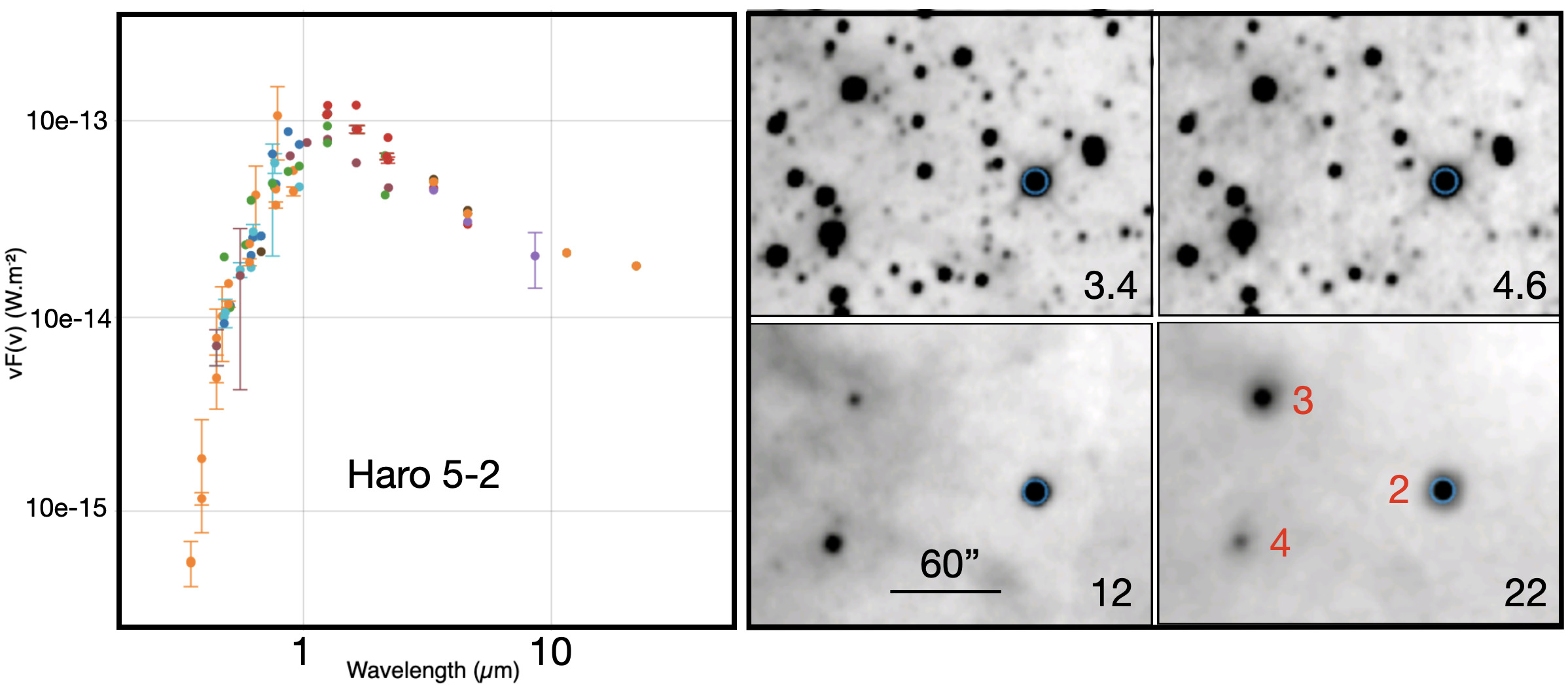}}
\caption{(left) Spectral energy distribution from SIMBAD of the integrated
light from the Haro~5-2 quadruple with data from SkyMapper, SDSS,
PanSTARRS, Johnson, 2MASS, AKARI, and WISE.  (right) WISE
images with Haro~5-2, -3, and -4 marked. The field is about 0.84~pc
wide at the distance of Haro~5-2.
\label{energydistribution}}
\end{figure*}

Haro~5-2 has been detected in a number of sky surveys from the
ultraviolet (Skymapper) to the mid-infrared
(WISE). Figure~\ref{energydistribution} shows the available photometry
together with four WISE panels that reveal the system to be bright at
infrared wavelengths. We have integrated over the energy distribution of the quadruple,
and derive a luminosity of 0.95~L$_\odot$ between 0.35~$\mu$m and
22~$\mu$m.  The peak of the energy distribution is around 1.05~$\mu$m, which
corresponds to about 2760~K. According to the temperature-spectral
type conversion by \cite{herczeg2014},  
this corresponds to an M7 spectral type. As discussed below this classification is
not borne out by spectroscopy. The energy distribution, although
dominated by the brightest member, Aa, is not from a single
object, but includes the three fainter components, thus shifting the peak 
in Figure~\ref{energydistribution} to
longer wavelengths. Also, Haro~5-2 shows H$\alpha$ emission, a sign
that accretion is occurring and indicating the presence of
circumstellar material that further adds to an infrared
excess. Thus, much of the light seen at longer wavelengths is not
likely to come from the dominant member Aa. In fact, almost half of
the observed luminosity, 0.43~L$_\odot$, falls within the four WISE
bands.

\subsection{Optical Spectroscopy}

\begin{figure*}[htb!]
\centering
\includegraphics[angle=0,width=17cm]{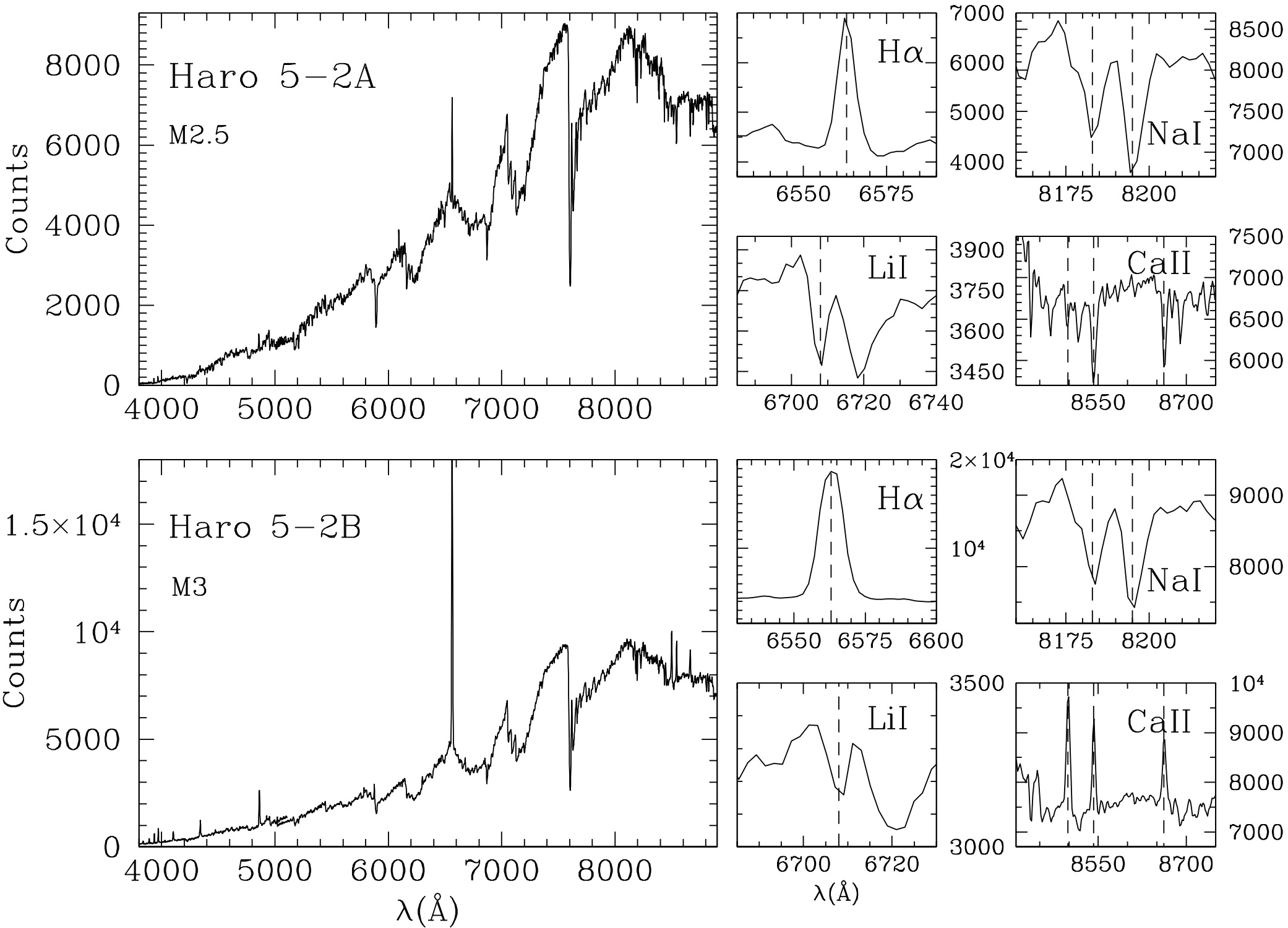}
\caption{Low resolution optical spectra of the Haro 5-2A and B components, 
 as obtained on UT2021-11-19 with the GHTS instrument at SOAR.
In the upper panels, the M2.5 WTTS Haro 5-2A component has H$\alpha$ weakly in emission.
In the lower panels,
the accreting M3 CTTS Haro 5-2B component has a strong H$\alpha$
emission line and the entire Balmer series is in emission.
Lithium is moderately strong. 
Ca H \& K are also found strongly in
emission,  as well as
the He I lines. Finally, the Ca II near-IR triplet is in emission. 
\label{soar_lowres}}
\end{figure*}

\begin{figure}[htb!]
\epsscale{0.9}
\centering
\centerline{\includegraphics[angle=0,width=10cm]{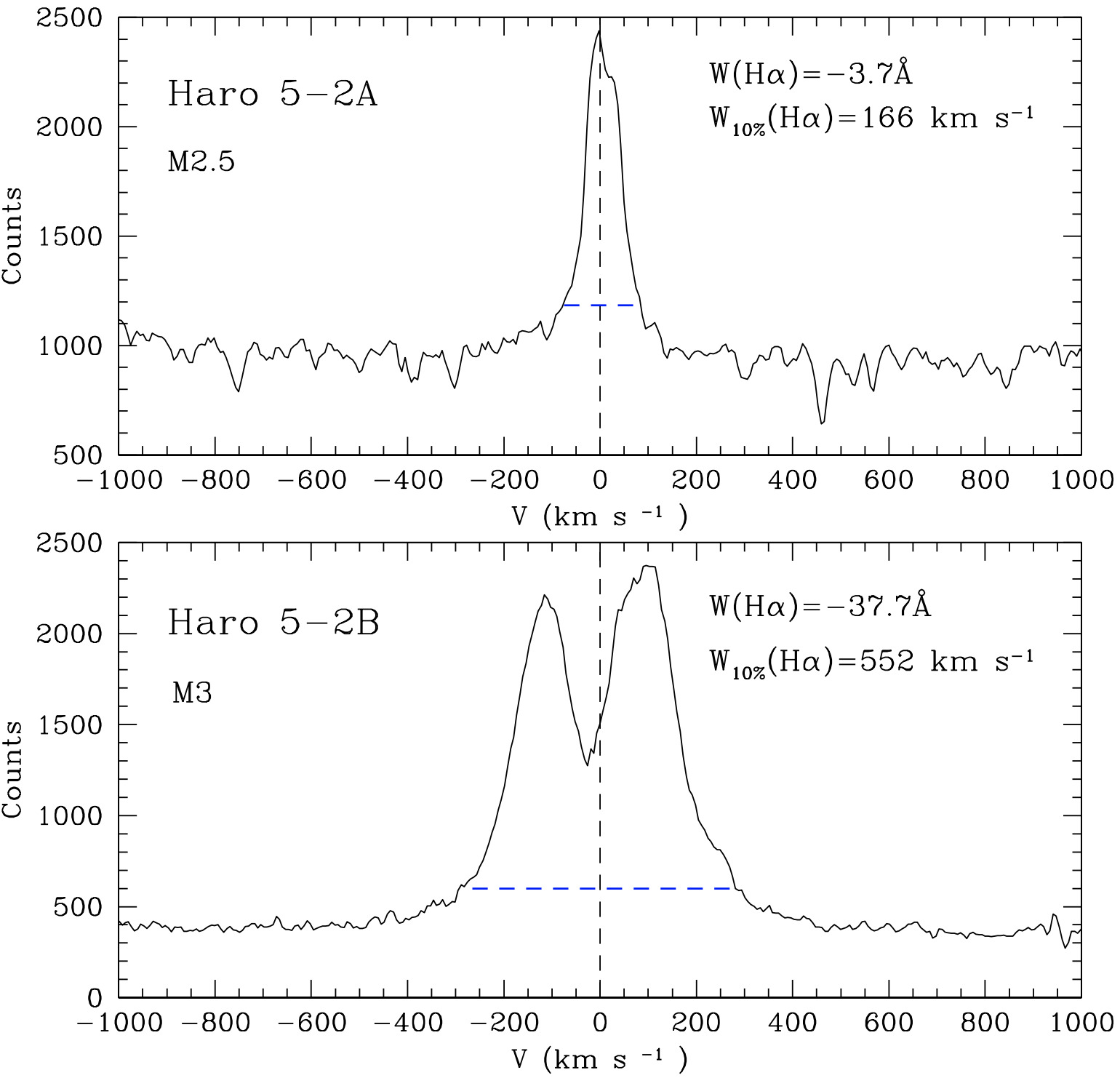}}
\caption{High resolution spectra of the H$\alpha$ line of the 
Haro 5-2 A and B components, obtained with the SOAR GHTS on UT 2021-11-19.
Upper panel: spectrum of the M2.5 WTTS Haro 5-2A.
Lower panel: spectrum of the accreting M3-type CTTS
Haro 5-2B.  The vertical dashed line is set at the rest wavelength for
H$\alpha$ ($6562.8${\AA}). The blue horizontal dashed line indicates the
Full Width of the H$\alpha$ profile at 10\% of its Full 
Height \citep{white2003}. 
\label{soar_hires}}
\end{figure}

\cite{haro1953} discovered  the T~Tauri nature of
Haro~5-2 on the basis of its H$\alpha$ emission, and it remains a
prominent H$\alpha$ emission line star
\citep{pettersson2014}. Figure~\ref{soar_lowres} shows optical
low-resolution spectra of the (unresolved) A and B
binaries, obtained with the GHTS instrument at SOAR (Section~2).

Both the A and B binaries have the TiO absorption bands
characteristic of M-type dwarfs. By comparing our spectra with spectra
of already known M-type T~Tauri stars (TTS) from the extensive sample
of \cite{briceno2019},
we assign spectral types of M2.5 for Haro 5-2A and M3 for Haro 5-2B, with
an overall uncertainty of 1 subclass.  As can also be seen in Figure
\ref{soar_lowres}, both the A and B binaries exhibit a clear Li I 6708{\AA }
absorption line, with equivalent widths W(Li I)$\sim 0.3${\AA }. Li I is a well known
indicator of youth for late type stars \citep[e.g.,][]{briceno1998,
briceno2001,white2003}.
Because lithium is depleted during the PMS stage in the deep convective
interiors of K and M-type stars, a late type star is classified as a
TTS if it has Li~I(6707\AA) in absorption, with an equivalent width
larger than that of a zero-age main sequence Pleiades star of the same
spectral type \citep{soderblom1993,garcia1994}. 
The Na I 8183,8195{\AA } absorption doublet is also a well known
feature useful for discriminating young, late-type PMS stars from
their field dwarf counterparts \citep{martin1996,lawson2009,lodieu2011,hillenbrand2013,hernandez2014,suarez2017, briceno2019}.
Consistent with the presence of Li I, both Haro 5-2A and B 
exhibit weak Na I absorption (Figure~\ref{soar_lowres}), with values
expected for TTS with ages of a few Myr, \citep[e.g., see Figure 13 in][]{briceno2019}.

\subsubsection{Haro 5-2A}

The low and higher resolution spectra show that the H$\alpha$ line of
the brighter A-binary exhibits a small equivalent width,
with an average value W(H$\alpha$)=-3.9{\AA}, and a range -3{\AA}
$\ge$ {W(H$\alpha$)} $\ge$ -4{\AA } during the $\sim$ 1.4 yr over
which we obtained multi-epoch spectroscopy (Tables
\ref{soar_spec_lowres} and \ref{soar_spec_hires}). In addition to the low
intensity of the emission, the H$\alpha$ line profile is narrow, $\la
200~\rm km \> s^{-1}$, as measured in the high-resolution
spectra. These characteristics are indicative of H$\alpha$ originating
in the active chromosphere of a young dwarf. The low intensity of
H$\alpha$ places the A-binary in the Weak-lined
T Tauri star class (see Figure~\ref{wha_spn}).

\subsubsection{Haro 5-2B}
\label{subsubsec:haro5-2b}
In contrast with its brighter sibling, Haro 5-2B shows a strong
H$\alpha$ emission line, which also is broad when seen in detail
in the higher resolution spectra. Moreover, thanks to our many
observing epochs (Tables \ref{soar_spec_lowres} and \ref{soar_spec_hires}), 
we find that the emission is highly variable,
ranging from W(H$\alpha$)=$-38.5${\AA } in November 2021 to a low 
W(H$\alpha$)=$-11.6${\AA } a year later, in November 2022.  We derive
an average W(H$\alpha$)=$-24.6${\AA }. 
The width of the H$\alpha$ line ranges from $\sim400$ to $>$500 km s$^{-1}$ 
and exhibits a very strong, slightly blueshifted absorption, 
suggestive of a strong wind. Moreover, the entire Balmer series is
found to be in emission (Figure~\ref{soar_lowres}), with $-3.6\la
$W(H$\beta$)$\la -12.3${\AA }. The Ca H \& K 3934,3968{\AA } lines are
also in emission, with equivalent widths in the range W(Ca H \& K)$\sim -3$ to $\sim -15${\AA}.
Both He I 5876{\AA } and He I 6678{\AA } are  also found in emission, 
with W(He I 5876) $\sim -2.1${\AA } and W(He I 6678)$\sim -1.2${\AA}. 
The Ca II near-IR triplet goes from being weakly in absorption
during the October 2021 observation, when Haro 5-2B was in a seemingly more
quiescent state, to being in emission, and staying in this state, 
from November 2021 to January 2023. Unfortunately we could not obtain spectra
with the reddest 400M2 configuration during the March 2023 observation.
We determine average values of 
W(CaII 8498)$=-1.7$\AA, W(Ca II 8542)$=-1.1$\AA, W(Ca II 8662)$=-1.0$\AA.

\begin{figure}[htb!]
\epsscale{1.0}
\centering
\centerline{\includegraphics[angle=0,width=8.5cm]{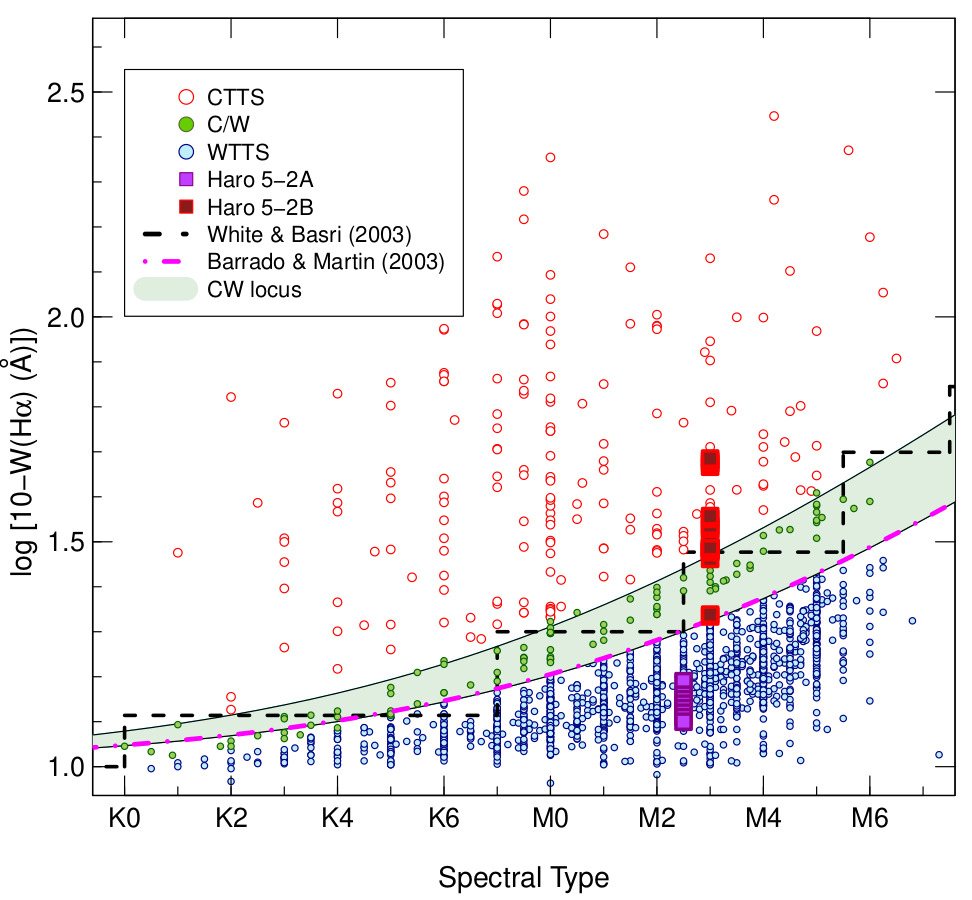}}
\caption{Equivalent width of H$\alpha$ (plotted 
as $\rm log_{10}(10-W(H\alpha)$) as a function of spectral type for Haro 5-2 A and B, at each epoch for which we obtained low resolution spectra at SOAR.  Haro 5-2A is plotted as large purple boxes, and Haro 5-2B as large red boxes. 
For comparison the Orion TTS from \cite{briceno2019} are also shown: 
CTTS as red circles, and WTTS as smaller blue dots. The
separation between CTTS and WTTS as defined by \cite{white2003} is shown
with the black dashed line, while the criterion adopted by
\cite{barrado2011} is plotted as a dash-dot magenta line. The
transition region adopted by \cite{briceno2019} between CTTS and WTTS
is shown by the pale green region where C/W objects fall (shown as
green dots). 
\label{wha_spn}}
\end{figure}

\begin{figure}[htb!]
\epsscale{1.0}
\centering
\centerline{\includegraphics[angle=0,width=8.5cm]{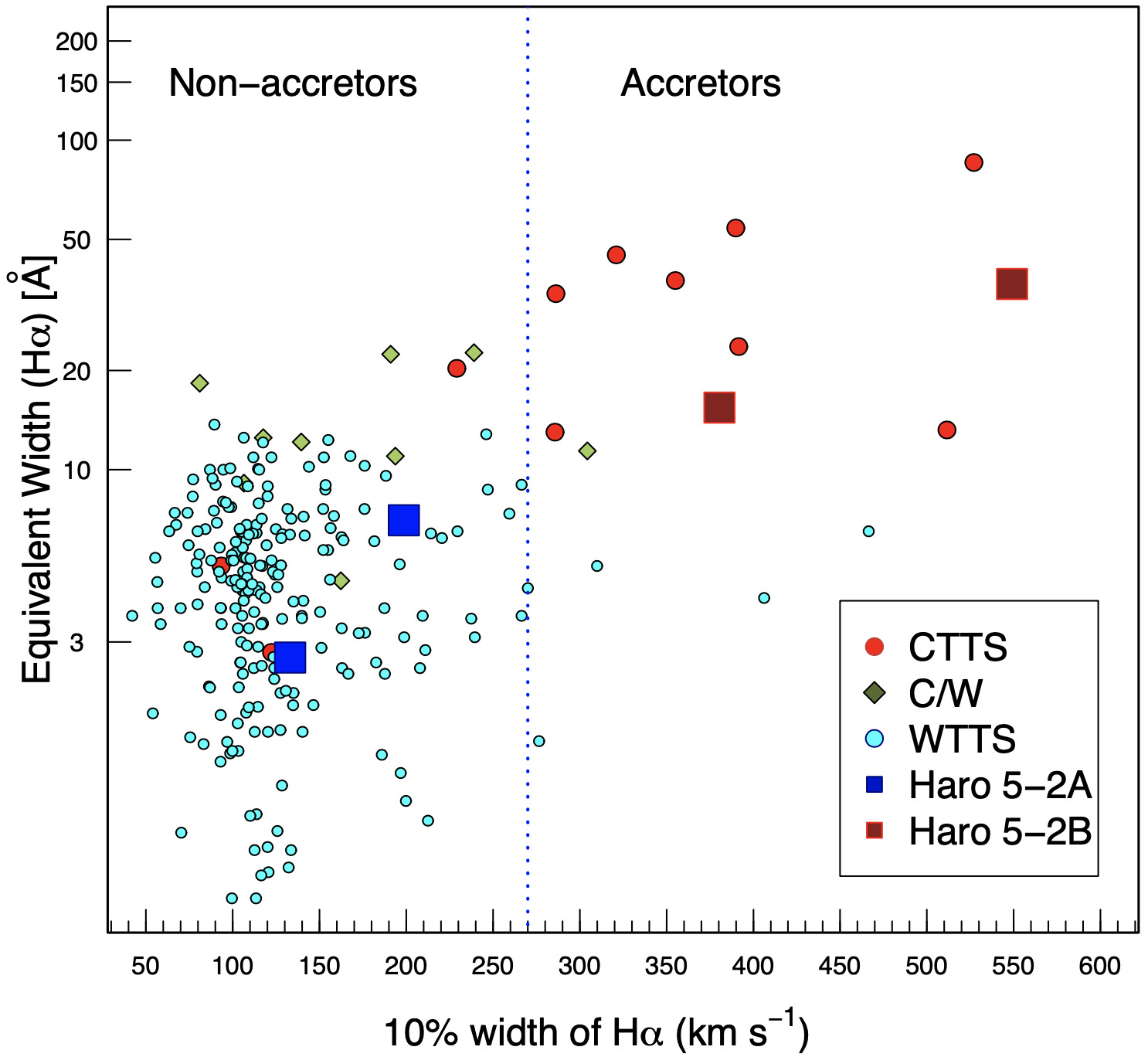}}
\caption{Equivalent width of H$\alpha$ as a function of the 10\% width
of the H$\alpha$ line profile, for Haro 5-2A (large blue boxes) and
Haro 5-2B (large red boxes), measured on the two epochs we obtained
high resolution spectra at SOAR.  Data are also shown for the Orion
TTS from \cite{briceno2019}: CTTS as red dots, WTTS as smaller cyan
dots, and C/Ws as green diamonds. The separation between CTTS and WTTS
as defined by \cite{white2003} is shown with the vertical blue dashed
line at $270\>\rm km \> s^{-1}$. 
\label{wha_w10}}
\end{figure}

The rich assortment of H, He, and Ca emission lines seen in Haro 5-2B,
together with the strength of the H$\alpha$ emission, the several
hundred $\rm km\> s^{-1}$ broad wings in its line profile and the
strong, slightly blueshifted central absorption, are all indicative of
a strongly accreting Classical T~Tauri star (CTTS), as can be seen by the location of
this object in the diagnostic diagrams of Figure \ref{wha_spn} and
Figure \ref{wha_w10}. The blueshifted central absorption is seen with
similar strength in both the October and November observations.

We highlight the importance of obtaining multi-epoch
spectra and of combining low- and higher-resolution spectra, in order to
provide both the wide spectral coverage needed to derive a reliable
spectral type ($\rm T_{eff}$), and the resolving power to characterize the
H$\alpha$ line profile, thus allowing to correctly classify young
PMS stars such as Haro 5-2. If we observed this object only in November, 
we would have misclassified Haro 5-2B as a Weak Line T~Tauri Star 
(WTTS) or C/W (Figure \ref{wha_spn}). 
It is the multi-epoch low resolution spectroscopy, combined with the higher 
resolution spectra, that allows us to confirm that the B
component is a CTTS, and that A is a WTTS. In particular, the
higher resolution spectra show that despite variations in both the
strength of H$\alpha$ and the velocity width, at both epochs Haro 5-2A
remains well inside the region populated by non-accreting,
young PMS low-mass stars, whereas the B-binary is well within 
the region where strongly accreting young PMS stars are found.

\subsubsection{Masses}

The optical spectra of Haro~5A (M2.5) and 5B (M3) are dominated by the components 
Haro~5Aa and 5Ba, respectively. If we assume a temperature uncertainty of 0.5 
sub-types we find that T$_{eff}$(Aa) $\sim$ 3485$\pm$80~K and T$_{eff}$(Ba) 
$\sim$ 3410$\pm$100~K using the spectral type to temperature conversion of 
\cite{herczeg2014}.  For an assumed age of 3~Myrs the magnetic models of 
\cite{feiden2016} suggest that Aa has a mass of 0.54$\pm$0.08~M$_{\odot}$ and 
Ba has a mass of 0.46$\pm$0.1~M$_{\odot}$. However, \cite{braun2021} and 
\cite{flores2022} have shown that magnetic models overpredict masses for young stars 
with M$_*$$<$0.4~M$_{\odot}$.  Non-magnetic Feiden models give   
M$_{Aa}$ = 0.35$\pm$0.05~M$_\odot$ and M$_{Ba}$ = 0.31$\pm$0.08~M$_\odot$. 
Given the uncertainties in both spectral types and in models, we end up with the not 
very accurate estimate of M$_{Aa}$ $\sim$ 0.45$\pm$0.15~M$_\odot$ and 
M$_{Ba}$ $\sim$ 0.35$\pm$0.15~M$_\odot$. The Ab and Bb components are likely to have 
somewhat lower masses. A rough estimate of the system mass is therefore $\sim$1.5~M$_\odot$. 

\subsection{Infrared Spectroscopy}

\begin{figure}[htb!]
\epsscale{0.5}
\centering
\centerline{\includegraphics[angle=0,width=8cm]{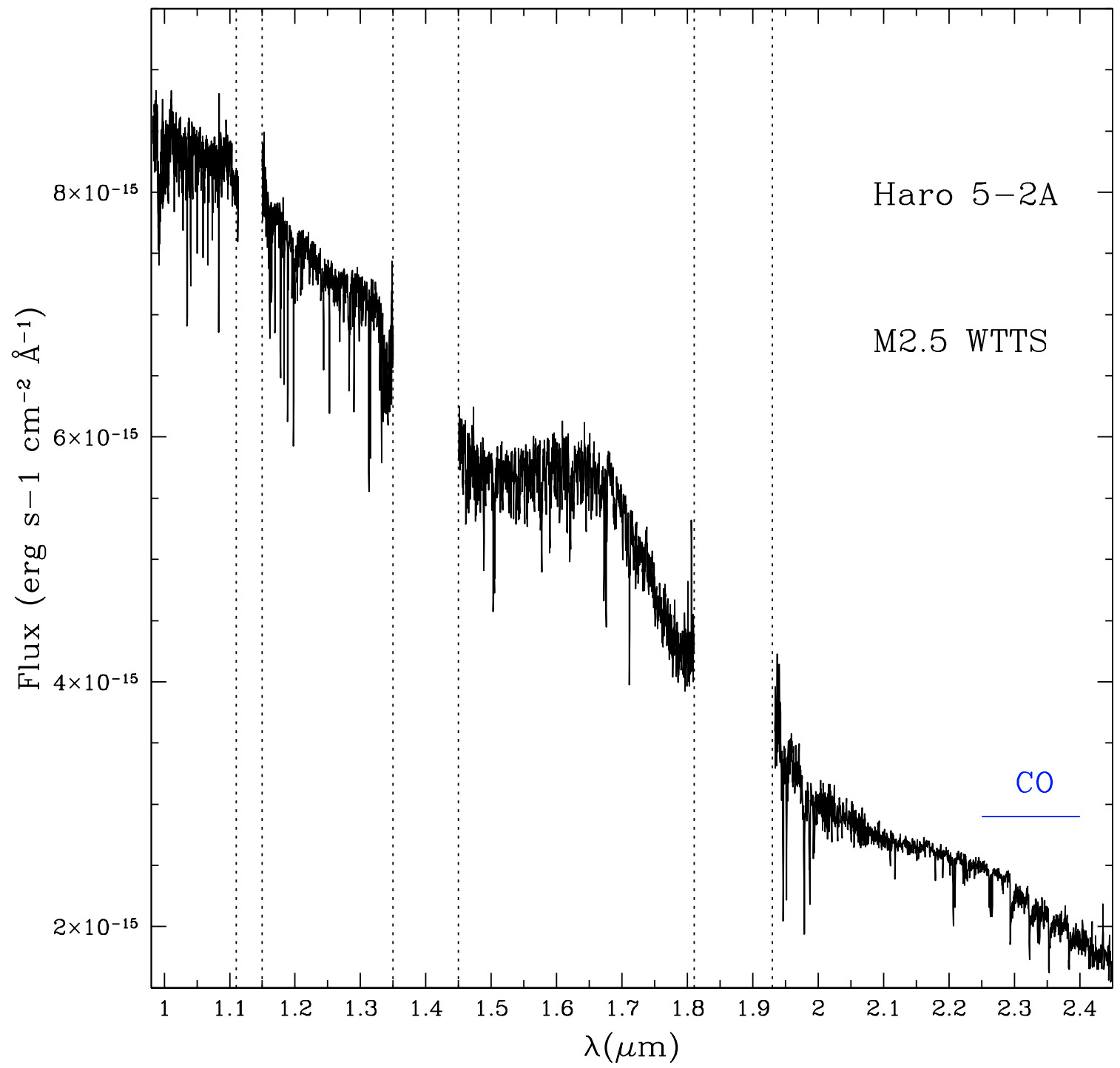}}
\centerline{\includegraphics[angle=0,width=8cm]{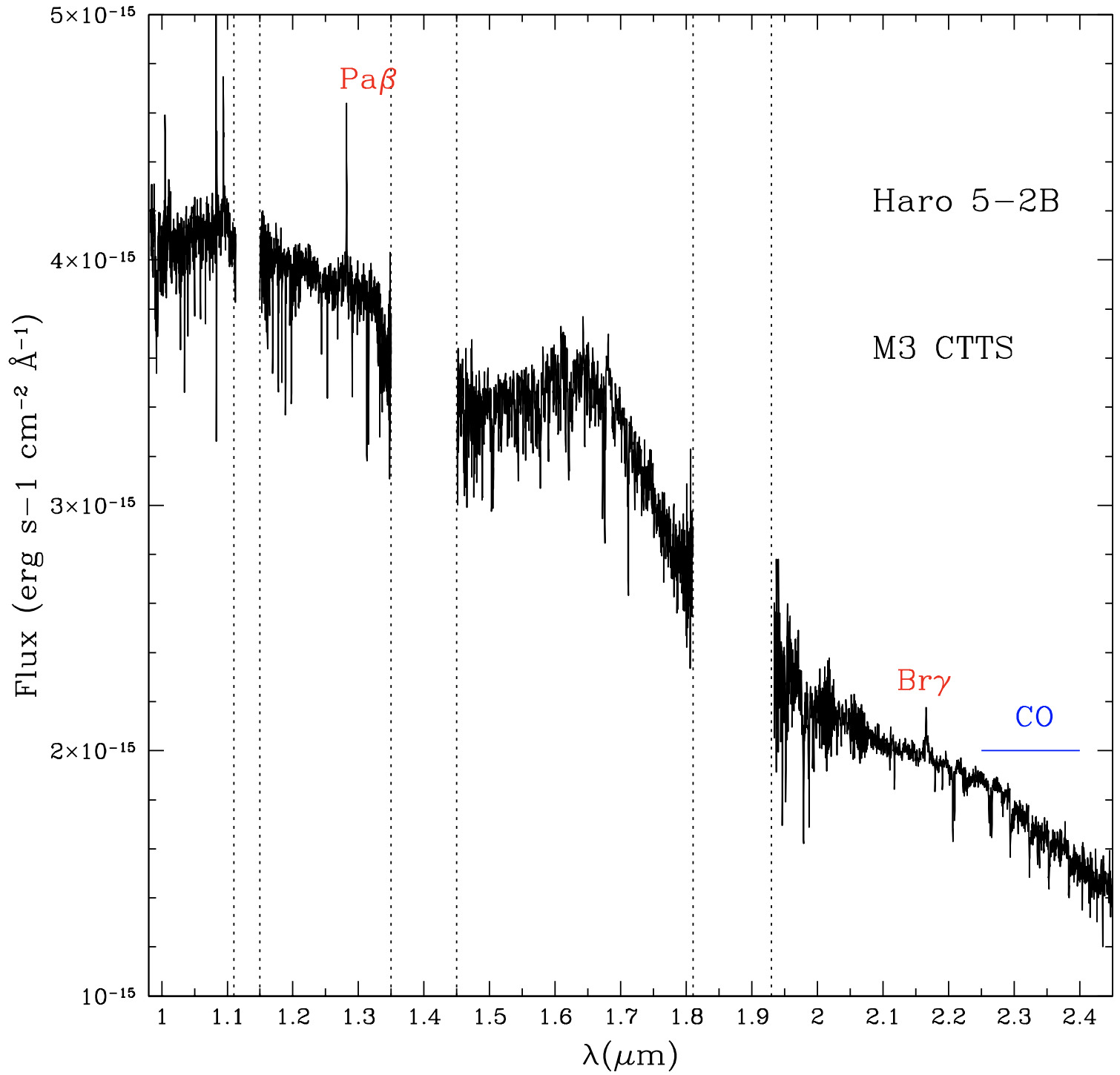}}
\caption{Near-IR spectra of Haro 5-2A and Haro 5-2B obtained with TSpec at the SOAR telescope.
Note the Pa$\beta$ and Br$\gamma$ emission lines in Haro~5-2B. Wavelength intervals with strong terrestrial opacity are excluded. 
\label{tspec_Haro_5-2}}
\end{figure}

The near-infrared spectra of the Haro~5-2A and 5-2B
(unresolved) binaries obtained with the TripleSpec 4.1 near-IR
spectrograph at the SOAR telescope are shown in
Figure~\ref{tspec_Haro_5-2}. Both objects display metal absorption
lines and CO bands and a triangular H-band continuum due to water
absorption, all consistent with their optical late spectral types. The
spectrum of Haro 5-2A (upper panel) shows no emission features,
consistent with its classification as a non-accreting WTTS, as shown
in Figures \ref{wha_spn} and \ref{wha_w10}. In contrast, the spectrum
of Haro~5-2B (lower panel) displays strong Paschen~$\beta$ and
Brackett~$\gamma$ emission lines, indicative of active accretion. The
near-infrared spectra, obtained on the same nights as the optical
spectra, are thus consistent with the optical classifications.

\begin{figure*}
\centerline{\includegraphics[angle=0,width=12cm]{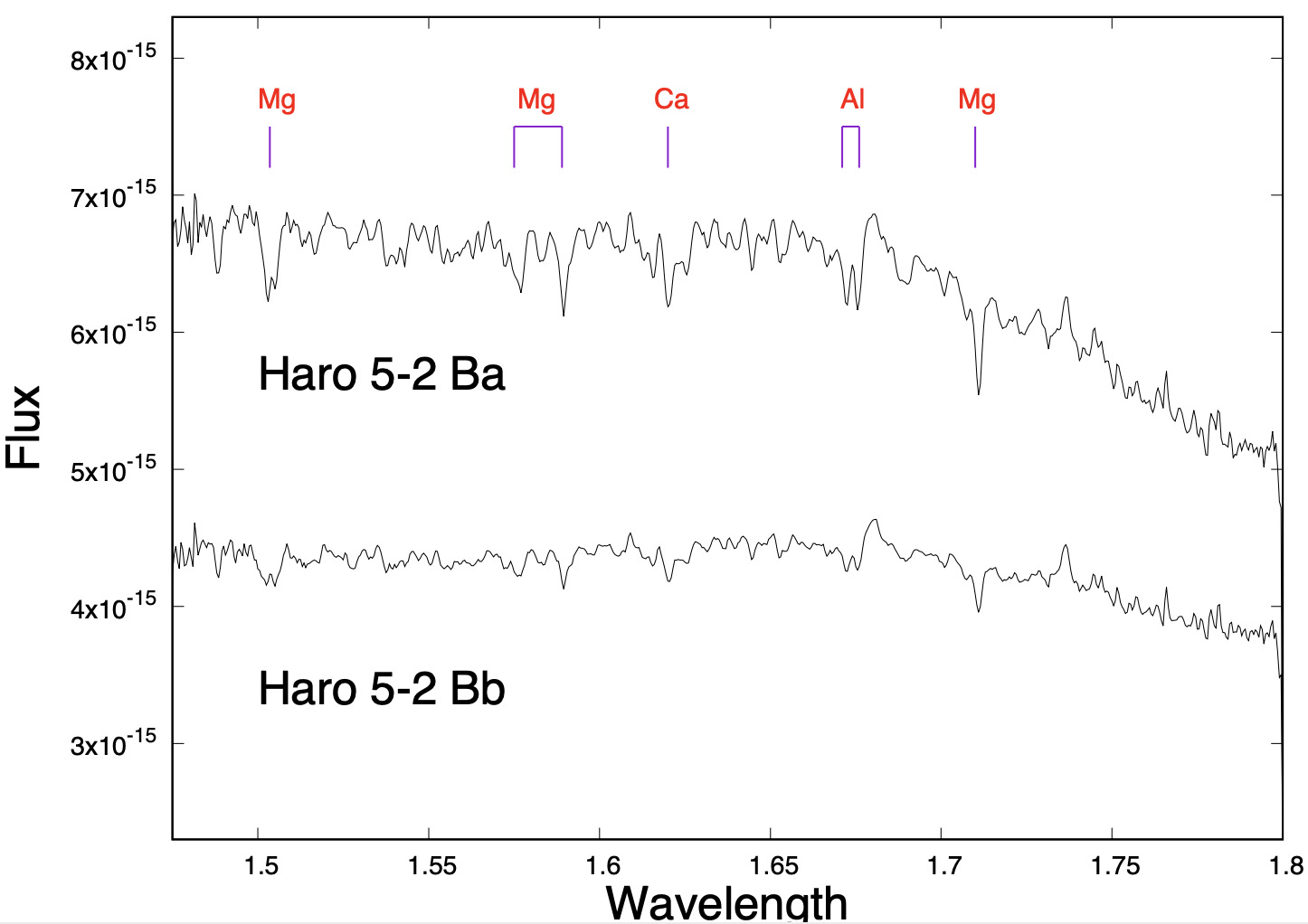}}
\caption{H-band spectra of the individual Ba and Bb
components resolved with adaptive optics and GNIRS at the Gemini-N 8m
telescope.  Wavelength is in microns and flux is in Wm$^{-2}$$\mu$m$^{-1}$.  
\label{gnirs-Ba-Bb}}
\end{figure*}

Line variability is always an issue when observing young stars, and we note
that a near-infrared spectrum of the (blended) A-binary obtained with
GNIRS on 2015 Sept 15 showed a broad Paschen~$\beta$ line in emission,
and a very weak emission at Brackett~$\gamma$, indicating that active
accretion was taking place at the time. On the same night, the
B-binary was also observed with GNIRS using adaptive optics, and the
two components Ba and Bb were resolved. The H-band spectra of the two
B-components are shown in Figure~\ref{gnirs-Ba-Bb}. The spectra are
almost identical, but the Bb component shows much more veiling as
well as a redder continuum. This indicates that the deep red color of
the Bb component seen in Figure~\ref{location}-bottom is not due to a
much later spectral type than Ba, but rather to a strong near-infrared
excess.

\section{Discussion}
\label{sec:discussion}



\subsection{Formation of Quadruple Systems}

As pointed out by \cite{larson1972}, multiple stars are a natural
consequence of the collapse of a rotating cloud core, and many if not
all stars may originate in such systems. Observationally it is well
established that in most star forming regions there is
an excess of young binaries and multiples relative to the field
population \citep[e.g.,][]{reipurth1993,leinert1993}. It follows that
to get down to the number in the field, there must be considerable
dynamical evolution during early stellar evolution, from N-body
interactions in combination with the presence of gas in the system.

The 2+1+1 quadruples are easily formed from small non-hierarchical
systems with 4 or more components through dynamical interactions which
lead to subsequent ejections of members into distant
bound orbits or into escapes \citep[e.g.,][]{delgado-donate2004}.
The more common 2+2 systems, on the other hand, have been explained in
a variety of ways.

\cite{bodenheimer1978} suggested that successive fragmentations of a
cloud core with transfer of spin angular momentum at each stage
would lead to wide binary systems in which each component is a close
binary. Such a cascade would naturally lead to 2+2 quadruples with
large ratios between inner and outer orbits, although hydrodynamical
simulations generally show more chaotic and sequential star
formation. The inner binaries of Haro 5-2 are well above the opacity 
limit to fragmentation, so this scenario might describe the formation of Haro~5-2.

\cite{whitworth2001} considered a different type of cascade in which two
colliding clouds form a shock-compressed layer that fragments into
filaments and cores and eventually form stars with massive
disks. Collisions of two disks around binaries can form bound
quadruple systems.

While turbulent core fragmentation will produce wider binaries and
multiples (100 - 10,000 AU), simulations of this process fail to
produce the very close binaries often found in 2+2 quadruples. To
produce such close binaries a dissipative gaseous environment is
required in which orbital decay will harden a binary
\citep[e.g.,][]{bate2002,delgado-donate2004,
lee2019,guszjenov2023,kuruwita2023}.
  
In disk fragmentation models, massive disks can become gravitationally
unstable and produce one or more companions and, combined with
capture, 2+2 systems can be formed \citep{kratter2016}. Filament
fragmentation has also been discussed as leading to bound binaries and
multiples \citep[e.g.,][]{bonnell1993,sadavoy2017,pineda2015}.

\cite{vanalbada1968a,vanalbada1968b} carried out N-body
simulations of small-N groups (10 to 24 stars), treating the stellar
components as point sources, and noting that among the final outcomes
were a number of 2+2 quadruples. The dynamical evolution of such gas free quadruple
systems has subsequently been further studied by many other authors, e.g.,  
\cite{harrington1974, mikkola1983,mikkola1984a,mikkola1984b,sterzik1998}.

To summarize, it appears that there are multiple viable pathways to
the formation of 2+2 quadruple stellar systems, and the multiple
systems we observe in the field may result from a mix of different
formation scenarios.

\subsection{The Origin of the Haro 5-2 Quadruple System}

In the following we discuss  three questions regarding
the structure, stability, and formation of the Haro 5-2 system.\\

{\em 1) Is Haro 5-2 merely a chance alignment of two young binaries
along the line-of-sight?}

We have photometric and spectroscopic evidence that the components of
Haro 5-2 are young, so we look at the surface density of nearby young
stars. In an area of roughly 25$\times$30 arcmin centered on Haro~5-2
there is a small group of only 9 additional YSOs. This immediately
indicates that the probability that Haro~5-2A and Haro~5-2B are close
due to a chance alignment along the line of sight is negligible.\\

{\em 2) Is Haro 5-2 a temporary quadruple or can it be in a stable
configuration?}

Compared to other pre-main sequence quadruples the most remarkable 
aspect of Haro~5-2 is the large separations of the two inner binaries relative
to the outer binary when compared to other known young 2+2 quadruple 
systems.\footnote{The ratio of projected outer and inner separations 
is 7.4. However, the boundary between chaotic and regular orbits is dependent on
many parameters, not least the eccentricity \citep{valtonen2006}, and a 
single value cannot grasp this complexity. But at least, for comparison, 
the ratio is 7.3 for HIP 28442, and since the star 
is a member of the old thick-disk population it has evidently been 
stable for a long time \citep{tokovinin2020}.} This opens the question 
of whether Haro 5-2 in the future might break apart.

For a definite answer one would need to know the orbits of the Haro
5-2 components, or at least the physical separations in the system,
the stellar masses, and the velocities of the components.  None of
this is currently known. Instead we have performed some statistical
estimates for a variety of system properties. For these calculations
we assumed that Aa has a mass of about 0.45~M$_{\odot}$ and the three
other components have masses of about 0.35~M$_{\odot}$, in total
1.5~M$_{\odot}$, which determines the gravitational potential. We
further assumed that the centers of mass of the two
subsystems are in the plane of the sky, since this would be the most
unstable configuration, while the orbital planes of the subsystems are
random. Finally we adopt random velocities assuming virial balance.
We then ran a code, described in more detail in \cite{reipurth2015},
which computes semi-major axes, eccentricities, and other orbital
characteristics for bound systems. After 10,000 experiments, about 1/3
remained stable 2+2 quadruples. Given that Haro 5-2 almost certainly
is extended along the line-of-sight and thus more stable, we conclude
that the quadruple may well be in a longterm stable configuration.\\

{\em 3) Could Haro 5-2 have become a quadruple at a later stage after
dispersal of most of the placental gas?}

\begin{figure*} 
\centerline{\includegraphics[angle=0,width=14cm]{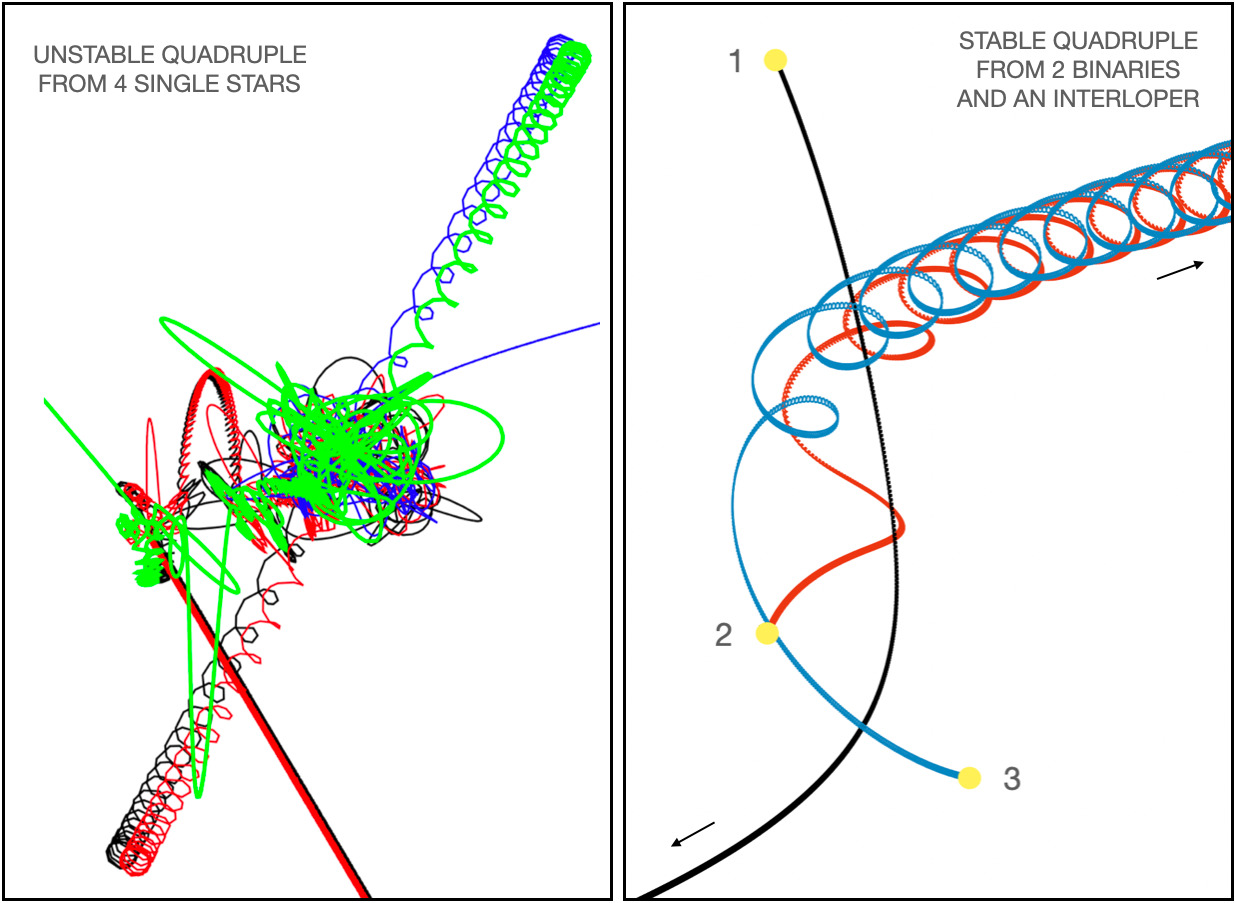}}
\caption{{\em (left)} 
Example of chaotic transformation of a non-hierarchical system of four
single stars into a temporary 2+2 quadruple(black/red + blue/green),
which subsequently breaks up into a binary (black/red) and two single
stars (blue and green). {\em (right)} Example of a non-hierarchical
configuration of two primordial binaries (2 and 3) that bind to form a
stable 2+2 quadruple system as an interloper flies by and carries away
the excess energy.  Formation of a quadruple system is far
simpler when the binaries have formed prior to their interaction.
Both simulations show the dynamical interactions of point sources
without any gas.
\label{seppo}}
\end{figure*}

If Haro 5-2 formed as a quadruple already during the embedded collapse
phase, one would expect that the resulting dynamical evolution in a
gaseous environment and active accretion would lead to a decrease of
the semi-major axes of the inner binaries, thus producing harder inner
binaries \citep[e.g.,][]{bate2002}. This is not what is seen in Haro
5-2, which has well resolved inner binaries, suggesting that it might
have developed its 2+2 configuration sufficiently late to 
have avoided early shrinking of the two inner binaries during the gas
rich embedded phase. This could have resulted in two spectroscopic
binaries as in the young 2+2 quadruples LkCa~3, EPIC 203868608, and
TIC 278956474 mentioned earlier. Consequently we have explored
numerically the dynamical evolution of a small cluster of single and
binary stars at a later stage, after the bulk of the gas has dispersed
or been accreted, to see if N-body simulations without gas could
result in a quadruple like Haro 5-2, following the pioneering work of
\cite{vanalbada1968a,vanalbada1968b}. We have performed many thousands of 
new numerical simulations with 4-, 10-, 20-, and 50-body groups of
single stars and masses drawn from the IMF and with virialized
velocities
\citep[see][for details of the code]{reipurth2010,reipurth2015}.   
Some binary and multiple systems indeed do form, including at least
temporarily some 2+2 quadruples (see Figure~\ref{seppo}a).

However, if one assumes that a substantial fraction of the bodies are
binaries then, not surprisingly, 2+2 systems are formed more
frequently (see Figure~\ref{seppo}b). We conclude that the number of
individual members of the initial multiple system is much less
important than whether some of those members are binaries.


Such 2+2 quadruples can form without the help of a gaseous environment 
when three objects, of which at least two are binaries, interact
chaotically,  in the process binding the two binaries together
into a 2+2 quadruple system, leaving the third body (single or binary)
to carry away the excess energy.
It should be noted that the binaries survive these interactions only
if they have binding energies that exceed the gravitational
perturbations induced by flybys. Effectively the binaries have to act
almost as point sources, that is, they must be close relative to the
impact parameter.  Thus, the stable 2+2 quadruples formed this way are
highly hierarchical, like $\epsilon$~Lyrae, because wider 
subsystems would be disrupted during the chaotic formation
process. It follows that to form 2+2 quadruples this way the two binaries must
previously have undergone a process that makes them ``hard''. This,
however, would not form the more open architecture of Haro 5-2.

Haro~5-2 is a member of the Ori~OB1b association, and is therefore not
a case of isolated star formation. Within the previously mentioned
little group of young stars surrounding Haro 5-2, the two closests are the
H$\alpha$ emitters Haro~5-3 and Haro~5-4 (see
Figure~\ref{energydistribution}), which are both only about 3.6~arcmin
away. This corresponds to about 80,000~AU or 0.4~pc in projection
which -- assuming a stellar velocity dispersion of $\sim$1~km~s$^{-1}$
-- can be traversed in less than half a million years. 
Haro~5-4 has a faint companion to the NNW which has an infrared excess
to 12~$\mu$m and possibly to 22~$\mu$m as seen in WISE images, and thus 
could be a young star, too.  It is therefore conceivable that the Haro~5-2
system in the past could have been part of a larger group that broke up,
with Haro~5-3 and 5-4 carrying away the energy that enabled the
Haro~5-2A and B binaries to bind together into a 2+2 system.

\clearpage

\section{Conclusions}
\label{sec:conclusions}

1. We have discovered a new young 2+2 quadruple system, Haro~5-2, in
   the 3-6~Myr old Ori~OB1b association, which encompasses the
   $\sigma$~Ori and the Collinder~70 clusterings. The system has an
   overall extent of 2.6$''$ and projected
   separations of the inner binaries of 0$\farcs$19 and 0$\farcs$35.

2. We obtained low-resolution optical spectra of both 
   the A and B components of the Haro 5-2 quadruple system on five 
   nights over $\sim$1.5 years, as well as two high-resolution spectra 
   separated by about a month.  The presence of significant
   TiO absorption molecular bands in the spectra, combined with
   indicators of stellar youth such as the Li I 6708~{\AA } and Na I
   8183,8195~{\AA} absorption features, confirm these as low-mass,
   young PMS stars. The brighter A component is a
   non-accreting M2.5 type ($T_{eff}\sim 3450 \deg K$) WTTS. The
   fainter B component is an M3 type ($T_{eff}\sim 3400 \deg K$)
   accreting CTTS, but on two nights it fell in the transition region between CTTS and WTTS. 
   Overall, the emission in both stars varies by up to a factor
   $2\times$ over the course of our observations.
   Assuming evolutionary models such as \cite{feiden2016}, 
   these spectral types correspond to a mass $\sim 0.35 - 0.45\,
   \msun$.  The H$\alpha$
   profile of the accreting B component shows broad wings extending to
   $\sim \pm 300\rm ~km\> s^{-1}$, and a strong, slightly blueshifted
   central absorption. This blueshifted absorption is seen in the
   higher resolution spectra at both epochs.  The Bb component is
   very red and the spectrum may be affected by accretion and by
   extinction from a circumstellar disk.



3. The hierarchy of Haro~5-2 is low, i.e. the two inner
   binaries in Haro 5-2 are unusually wide relative to the separation 
   between the two binaries in comparison with oter known PMS quadruples. 
   We have made simulations of a variety of
   configurations and conclude that the system may well survive
   intact over long time scales.

4. Since the members of the Haro~5-2 quadruple are well resolved and
   presumably of the same age, this system may provide stringent
   constraints on PMS evolutionary models, similar to the well-studied
   pentuple GG~Tau.  Haro~5-2 also displays a significant infrared
   excess, and ALMA observations of its circumstellar material may
   offer insights into the effect of flybys on disks.

\acknowledgments

{\bf Acknowledgements}

We thank A. Tokovinin and J.\,A. Caballero for valuable comments, 
and A. Tokovinin for providing the data in Table~\ref{tokovinin} .

Based in part on observations obtained at the International Gemini Observatory, 
a program of NSF's NOIRLAB, which is managed by  the Association of Universities for Research in Astronomy (AURA),
under a cooperative agreement with the National Science Foundation on behalf of the
Gemini partnership: the National Science Foundation (United States),
the National Research Council (Canada), Agencia Nacional de Investigaci\'on y Desarollo (Chile),
Ministerio de Ciencia, Tecnologia e Innovaci\'on (Argentina), and
Ministerio da Ciencia, Technologia, Inovacoes e Comunicacoes (Brazil), 
and Korea Astronomy and Space Science Institute (Republic of Korea)..

It is also based in part on observations obtained at the Southern Astrophysical
Research (SOAR) telescope, which is a joint project of the
Minist\'{e}rio da Ci\^{e}ncia, Tecnologia e Inova\c{c}\~{o}es
(MCTI/LNA) do Brasil, the US National Science Foundation's NOIRLab,
the University of North Carolina at Chapel Hill (UNC), and Michigan
State University (MSU).  IRAF was distributed by the National Optical
Astronomy Observatory, which was managed by the Association of
Universities for Research in Astronomy (AURA) under a cooperative
agreement with the National Science Foundation. 

The Robo-AO system is supported by collaborating partner institutions,
the California Institute of Technology and the Inter-University Centre
for Astronomy and Astrophysics, by the National Science Foundation
under Grant Nos. AST-0906060, AST-0960343, and AST1207891, by a grant
from the Mt. Cuba Astronomical Foundation and by a gift from Samuel
Oschin.

This paper includes data collected by the TESS mission. Funding for the 
TESS mission is provided by the NASA's Science Mission Directorate.


This research has made use of the SIMBAD database, operated at CDS,
Strasbourg, France, and of NASA's Astrophysics Data System
Bibliographic Services.

This research made
use of Astropy, a community-developed core Python package for
Astronomy \citep{astropy2013,astropy2018,astropy2022}.  We are grateful to
Katelyn Allers for her detailed and instructive YouTube videos on
reducing TSpec data:\\
https://www.youtube.com/user/katelynallers/videos.

\software{Astropy, IRAF, TOPCAT (Taylor 2005), Lightkurve (LightKurve Collaboration,  2018)}

{\it Facilities:}  \facility{PO:1.5m (Robo-AO), Keck:II (NIRC2-LGS),
Gemini-North (GNIRS,NIRI), SOAR (Goodman spectrograph, TSpec spectrograph, HRCam)}

\clearpage


\clearpage

\clearpage

\begin{deluxetable}{cccccl}[ht!]
\tabletypesize{\scriptsize}
\tablewidth{0pt}
\tablecaption{SOAR Goodman GHTS Observations of Haro 5-2\label{soar_ghts_obs}}
\tablehead{
\colhead{Component} & \colhead{UT Date} & \colhead{Setup} & \colhead{Slit} & \colhead{Binning} & \colhead{Exposures} \\ 
\colhead{} & \colhead{(yyyy-mm-dd)} & \colhead{} & \colhead{($\arcsec$)} & \colhead{} & \colhead{($N \times s$)}
}
\startdata
A & 2021-10-20  &  400M1 & 1.0 & $2\times 2$ & $3\times 120$ \\
A & 2021-10-20  &  400M2 & 1.0 & $2\times 2$ & $3\times 120$ \\
B & 2021-10-20  &  400M1 & 1.0 & $2\times 2$ & $3\times 120$ \\
B & 2021-10-20  &  400M2 & Brice1.0 & $2\times 2$ & $3\times 120$ \\
B & 2021-10-20  &  400M1 & 1.0 & $2\times 2$ & $3\times 300$ \\
B & 2021-10-20  &  400M2 & 1.0 & $2\times 2$ & $3\times 300$ \\
A & 2021-10-21  &  $2100\_650$ & 0.45 & $1\times 2$ & $3\times 900$ \\
B & 2021-10-21  &  $2100\_650$ & 0.45 & $1\times 2$ & $3\times 1200$ \\
A & 2021-11-19  &  400M1 & 1.0 & $2\times 2$ & $3\times 120$ \\
A & 2021-11-19  &  400M2 & 1.0 & $2\times 2$ & $3\times 120$ \\
B & 2021-11-19  &  400M1 & 1.0 & $2\times 2$ & $3\times 300$ \\
B & 2021-11-19  &  400M2 & 1.0 & $2\times 2$ & $3\times 300$ \\
A & 2021-11-19  &  $2100\_650$ & 0.45 & $1\times 2$ & $3\times 900$ \\
B & 2021-11-19  &  $2100\_650$ & 0.45 & $1\times 2$ & $3\times 1200$ \\
A & 2022-11-10  &  400M1 & 1.0 & $2\times 2$ & $3\times 120$ \\
B & 2022-11-10  &  400M1 & 1.0 & $2\times 2$ & $1\times 300$ \\
A & 2022-11-10  &  400M2 & 1.0 & $2\times 2$ & $1\times 120$ \\
B & 2022-11-10  &  400M2 & 1.0 & $2\times 2$ & $1\times 300$ \\
A & 2023-01-07  &  400M1 & 1.0 & $2\times 2$ & $1\times 200$ \\
B & 2023-01-07  &  400M1 & 1.0 & $2\times 2$ & $1\times 400$ \\
A & 2023-01-07  &  400M2 & 1.0 & $2\times 2$ & $1\times 200$ \\
B & 2023-01-07  &  400M2 & 1.0 & $2\times 2$ & $2\times 400$ \\
A & 2023-03-10  &  400M1 & 1.0 & $2\times 2$ & $3\times 200$ \\
B & 2023-03-10  &  400M1 & 1.0 & $2\times 2$ & $2\times 600$ \\
\enddata
\end{deluxetable}

\begin{deluxetable*}{llccccccccc}[ht!]
\tabletypesize{\scriptsize}
\tablewidth{0pt}
\tablecaption{Equivalent Widths in SOAR low-resolution Spectra of Haro 5-2}
\tablehead{
\colhead{Component} & \colhead{Spectral} & \colhead{Type} & \colhead{UT DATE} & \colhead{W(H$\alpha$)} & \colhead{W(H$\beta$)} & \colhead{W(Li I)} & \colhead{W(NaI)} &  
\colhead{W(CaII 8498)} & \colhead{W(CaII 8542)} &\colhead{W(CaII 8662)} \\
\colhead{} & \colhead{Type} & \colhead{} & \colhead{yyyy-mm-dd} & \colhead{(\AA)} & \colhead{(\AA)} & \colhead{(\AA)} & \colhead{(\AA)} & \colhead{(\AA)} & \colhead{(\AA)} & \colhead{(\AA)}
}
\startdata
A  & M2.5 &  W & 2021-10-20 & $ -4.3\pm 0.1$ & $ -2.7\pm 0.1$ & $0.4\pm 0.1$ & $1.8\pm 0.1$ & $ 0.2\pm 0.1$ & $ 1.2\pm 0.1$ & $ 0.9\pm 0.1$ \\
A  & M2.5 &  W & 2021-11-19 & $ -4.1\pm 0.2$ & $ -2.6\pm 0.1$ & $0.4\pm 0.1$ & $1.7\pm 0.1$ & $ 0.2\pm 0.1$ & $ 1.3\pm 0.1$ & $ 1.0\pm 0.1$ \\
A  & M2.5 &  W & 2022-11-10 & $ -3.7\pm 0.1$ & $ -3.0\pm 0.1$ & $0.2\pm 0.1$ & $1.2\pm 0.1$ & $ 0.3\pm 0.1$ & $ 0.5\pm 0.1$ & $ 0.6\pm 0.1$ \\
A  & M2.5 &  W & 2023-01-07 & $ -4.2\pm 0.2$ & $ -3.1\pm 0.1$ & $0.3\pm 0.1$ & $1.8\pm 0.1$ & $ 0.6\pm 0.1$ & $ 1.1\pm 0.1$ & $ 0.9\pm 0.1$ \\
A  & M2.5 &  W & 2023-03-10 & $ -3.3\pm 0.1$ & $ -1.8\pm 0.1$ & $0.2\pm 0.1$ & \nodata & \nodata & \nodata &  \nodata \\
B  & M3   &  C & 2021-10-20 & $-20.3\pm 0.3$ & $ -5.8\pm 1.3$ & $0.4\pm 0.1$ & $1.8\pm 0.1$ & $ 0.1\pm 0.1$ & $ 0.3\pm 0.2$ & $ 0.4\pm 0.1$ \\
B  & M3   &  C & 2021-11-19 & $-37.5\pm 0.8$ & $-12.0\pm 0.5$ & $0.3\pm 0.1$ & $1.7\pm 0.1$ & $-2.34\pm 0.1$ & $-1.6\pm 0.1$ & $-1.3\pm 0.1$ \\
B  & M3   &  C & 2022-11-10 & $-11.7\pm 0.1$ & $-4.3\pm 0.1$ & $0.3\pm 0.1$ & $1.2\pm 0.1$ & $-0.6\pm 0.1$ & $-0.3\pm 0.1$ & $-0.3\pm 0.1$ \\
B  & M3   &  C & 2023-01-07 & $-24.6\pm 0.7$ & $-5.6\pm 0.1$ & $0.3\pm 0.1$ & $1.5\pm 0.1$ & $-1.2\pm 0.1$ & $-0.9\pm 0.1$ & $-1.0\pm 0.1$ \\
B  & M3   &  C & 2023-03-10 & $-24.6\pm 2.0$ & $-6.9\pm 2.0$ & $0.2\pm 0.1$ & 
\nodata   & \nodata & \nodata & \nodata \\
\enddata
\tablenotetext{}{(1): C=Classical TTS; W=Weak-line TTS}
\tablenotetext{}{(2): For the Na I doublet we report the combined equivalent width of the 8183{\AA } and 8195{\AA } lines}
\label{soar_spec_lowres}
\end{deluxetable*}

\clearpage

\begin{deluxetable}{cccc}[ht!]
\tablewidth{0pt}
\tablecaption{H$\alpha$ Emission Line in SOAR high-resolution Spectra of Haro 5-2}
\tablehead{
\colhead{Component} & \colhead{UT DATE} & \colhead{W(H$\alpha$)} & \colhead{W10(H$\alpha$)} \\
\colhead{} & \colhead{(yyyy-mm-dd)} & \colhead{(\AA)} & \colhead{($km\> s^{-1}$)} 
}
\startdata
Haro 5-2A & 2021-10-21 & $ -7.08\pm 0.13$ &  $198.6\pm 3.6$ \\
Haro 5-2A & 2021-11-19 & $ -3.69\pm 0.03$ &  $165.7\pm 2.7$ \\
Haro 5-2B & 2021-10-21 & $-18.73\pm 0.24$ &  $420.8\pm 2.8$ \\
Haro 5-2B & 2021-11-19 & $-37.73\pm 0.62$ &  $552.2\pm 2.9$ \\
\enddata
\tablenotetext{}{}
\label{soar_spec_hires}
\end{deluxetable}

\begin{deluxetable}{ccccc}[ht!]
\tablewidth{0pt}
\tablecaption{Separation and Position Angles of Haro 5-2 Pairs}
\tablehead{
\colhead{Pair} & \colhead{PA}     & \colhead{Sep.}   & \colhead{Proj. Sep.} & \colhead{$\Delta$I} \\
\colhead{}     & \colhead{(deg.)} & \colhead{($''$)} & \colhead{(AU)}       & \colhead{(mag)} 
}    
\startdata
Aa,Ba &  124.9$\pm$0.8 & 2.6130$\pm$0.0008 & 975 & \\
Aa,Ab &   54.7$\pm$0.8 & 0.1942$\pm$0.0008 &  72 & 1.0 \\
Ba,Bb &  233.0$\pm$1.3 & 0.3504$\pm$0.0013 & 131 & 0.7 \\
\enddata
\tablenotetext{}{(1): Observed by Andrei Tokovinin
on 2021.7983 with the HRCam speckle camera at the SOAR telescope in good
seeing \citep{tokovinin2022}
}
\tablenotetext{}{(2): The data for the Ba,Bb pair are noisy.}
\tablenotetext{}{(3): Projected separations for a distance of 373~pc}
\label{tokovinin}
\end{deluxetable}

\end{document}